%% file: Main.tex
\documentclass[conference, final, twocolumn]{IEEEtran}
\usepackage{graphicx}

\usepackage[utf8]{inputenc}
\usepackage[english]{babel}
\usepackage{import}
\usepackage{calc,amsfonts,graphicx,cite,color,setspace,epsfig,amssymb, amsthm,tikz,graphics,dcolumn, array} 
\usepackage{multirow, float,epstopdf,paralist,lmodern}	
\usepackage{subcaption,balance,caption,subcaption,url,multicol,blindtext}
\usepackage[none]{hyphenat}
\usepackage{amsmath} 
\usepackage[normalem]{ulem}

\usepackage[normalem]{ulem}
\newtheoremstyle{mystyle}%              % Name
{}%                                     % Space above
{}%                                     % Space below
{\itshape}%                             % Body font
{}%                                     % Indent amount
{\it}%\bfseries                            % Theorem head font
{.}%                                    % Punctuation after theorem head
{ }%                                    % Space after theorem head, ' ', or \newline
{}%                                     % Theorem head spec (can be left empty, meaning `normal')
\theoremstyle{mystyle}
\newtheorem{thrm}{Theorem}
\newtheorem{prop}[thrm]{Proposition}
\newtheorem{lem}[thrm]{Lemma} 

\newtheorem{remark}{Remark}

\newtheorem{example}{Example}
 
\graphicspath{{./Figures/}}

\usepackage[english]{babel}
\usepackage[algo2e]{algorithm2e} 
\usepackage{algorithm}
\usepackage{algpseudocode}
\usepackage{filecontents}

\usepackage[font={footnotesize}]{caption}
\begin{document}

\title{Optimal Locally Repairable Codes with Improved Update Complexity}

\author{Mehrtash Mehrabi, \IEEEmembership{Student Member, IEEE,}
\and         Mostafa~Shahabinejad,~\IEEEmembership{Student~Member,~IEEE,}
\and~Masoud~Ardakani,~and~\IEEEmembership{Senior~Member,~IEEE,}
\and        Majid~Khabbazian~\IEEEmembership{Member,~IEEE,}
%\thanks{The authors are with the Department of Electrical and Computer Engineering, University of Alberta, Edmonton, AB, Canada (email: %\{mehrtash, mshahabi, mkhabbazian, ardakani\}@ualberta.ca).}
}

\maketitle

\renewcommand{\thefigure}{\arabic{figure}}
%\large

\begin{abstract}
For a systematic erasure code, update complexity (UC) is defined as the maximum number of parity blocks needed to be changed when some information blocks are updated. 
Locally repairable codes (LRCs) have been recently proposed and used in real-world distributed storage systems.
In this paper, update complexity of optimal LRCs is studied and both lower and upper bounds on UC are established in terms of length ($n$), dimension ($k$), minimum distance ($d$), and locality ($r$) of the code, when $(r+1)\mid n$.
%and $R>(1-\frac{1}{\sqrt{n}})^2$. 
Furthermore, a class of optimal LRCs with small UC is proposed.
Our proposed LRCs could be of interest as they improve UC without sacrificing optimality of the code. 
\end{abstract}


\IEEEpeerreviewmaketitle

\section{Introduction}\label{sec:Introduction}
\import{Sections/}{Introduction.tex}

\section{Preliminaries}\label{sec:preliminaries}
\import{Sections/}{Preliminaries.tex}

\section{Bound On Update Complexity (UC)}\label{sec:BoundOnUpdateComplexity(UC)}
\import{Sections/}{ProposedBound.tex}

\section{Our Proposed Optimal LRCs With Small UC}\label{sec:CodingMethodwithLowUC}
\import{Sections/}{ProposedLRC.tex}

%\section{Comparison With Other Coding Methods}\label{sec:Comparison}
%\import{Sections/}{Comparison.tex}

\section{Conclusion}\label{sec:Conclusion}
\import{Sections/}{Conclusion.tex}

\bibliographystyle{IEEEtran}
\bibliography{IEEEabrv,References}

\appendices
%\section{Proof of Proposition \ref{prop:rate}}
%\label{app:Proof_rate}
%\import{Sections/}{Proof_rate.tex}
%%\section{Proof of Lemma \ref{lem:parityLRC}}
%%\label{app:Proof_lemma1}
%%\import{Sections/}{Proof_lemma1.tex}
\section{Proof of Theorem \ref{thrm:UCbound}}
\label{app:Proof_UCbound}
\import{Sections/}{boundUC.tex}

\section{Proof of Theorem \ref{thrm:aveUC1}}
\label{app:Proof_UC1}
\import{Sections/}{Proof_UC1.tex}
\section{Proof of Proposition \ref{prop:min_d}}
\label{app:Proof_d}
\import{Sections/}{proof_d.tex}

\end{document}

%% file: Sections/Introduction.tex
%$\bullet$ Benefits of Distributed Storage Systems (DSS)
Distributed storage systems (DSSs) are used to store large-scale data in a secure and reliable way. 
A DSS uses a number of storage nodes, called data nodes, to store data. 
In such storage systems, data node failures occur frequently due to several reasons such as hardware/software problems associated with the underlying network or data nodes. 
In order to recover the lost/erased data, redundancy is required.
For example, the approach of keeping several replicas of data in distinct data nodes, known as replication, is widely used \cite{XORingElephants}.
The high storage overhead of the replication method results in a costly maintenance for DSSs \cite{WAS}. 

Recently, systematic erasure codes have been proposed and used in DSSs to decrease storage overhead. 
In order to use an erasure code in a DSS, first, a stripe of data is split into $k$ information blocks. Then, using an $(n,k)$ erasure code, $n$ encoded blocks are generated from $k$ information blocks. The $n$ encoded blocks are then stored in $n$ different data nodes. 
Hence, for an $(n,k)$ erasure code with minimum distance $d$, where $d \leq n-k+1$, $n-k$ parity blocks are generated such that $k$ information blocks can be recovered by any ~$n - d + 1$~ encoded blocks. 
Such, systematic linear block erasure codes\footnote{In systematic codes, information blocks can be directly stored and read with no encoding and decoding processes \cite{WAS,XORingElephants}. This is why, in DSSs, systematic codes are preferred to the non-systematic ones. From now on, whenever we say any class of erasure codes we mean the systematic erasure codes of that class.} have been used in real-world cloud storage systems such as Google File System \cite{Google_Availability}, Microsoft Azure Storage \cite{WAS}, and Facebook HDFS-RAID \cite{XORingElephants}.     
%$\bullet$ Existing Solutions (1. regenerating codes, 2. Codes with disk I/O) 3. LRCs

Although erasure codes reduce storage overhead, they need access to many nodes to recover a missing data block. 
For example, an $(n,k)$ erasure code with $d=n-k+1$, known as maximum distance separable (MDS) codes, requires $k$ data nodes in order to recover one missing data block.    

%require high repair bandwidth, disk I/O, and computation. Therefore, regenerating codes, low disk I/O codes, and LRC codes have been proposed, to reduce repair bandwidth, disk I/O, and code locality, respectively.
%Repair bandwidth is defined as the amount of download required to recover a failed node and is one of the important factors in erasure codes which should be reduced. Reducing repair bandwidth needs increasing storage overhead, and there is a bound between them, proposed in [] to choose the efficient values. Regenerating codes are class of codes which are formed based on reducing the repair bandwidth in which each block of data is divided to some sub-blocks and to reconstruct the failed data only some sub-blocks of other nodes need to be download and as a result the repair bandwidth will be decreased.   
%Disk I/O, which is number of reads required for reconstruction failed node, also considered as a criterion to evaluate the efficiency of code, so, many codes have been proposed to deal with it such as []. 
Decreasing the number of participating nodes in a recovery process is crucial as it decreases the costly repair bandwidth and disk I/O.
The maximum number of active nodes required to recover a failed node, denoted $r$, is defined as the locality of a erasure code. 
Locally repairable codes (LRCs) are a class of codes that are designed for small $r$.
The following bound is obtained for minimum distance of LRCs \cite{lrcHuang,lrcDimakis}
\begin{equation}\label{eq:minimumdistancebound} 
d \le n - k - \Big\lceil \frac{k}{r} \Big\rceil + 2.
\end{equation}
LRCs that achieve this bound are called \textit{optimal}.
It is verified that the bound in (\ref{eq:minimumdistancebound}) is tight if $(r+1)\mid n$ \cite{lrcDimakis}. 
From now on, we assume that $(r+1)\mid n$.
\begin{figure}
	\centering
	\begin{subfigure}[b]{0.33\textwidth}
		\includegraphics[width=\textwidth]{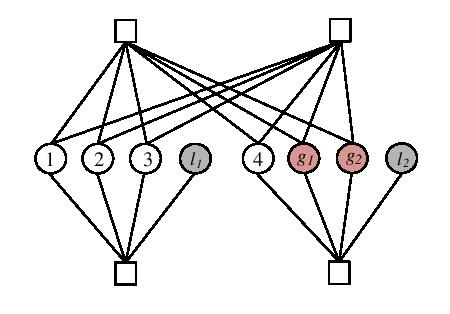}
		\caption{}
		\label{fig:1a}
	\end{subfigure}
	%\hspace{1.4cm}%add desired spacing between images, e. g. ~, \quad, \qquad, \hfill etc. 
	\begin{subfigure}[b]{0.33\textwidth}
		\includegraphics[width=\textwidth]{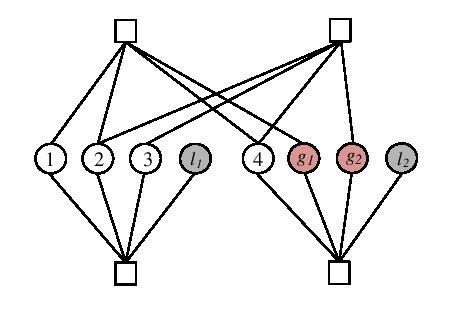}
		\caption{}
		\label{fig:1b}
	\end{subfigure}
	\caption{Tanner graphs of two optimal $(n,k,d,r)=(8,4,4,3)$ LRCs with different values of UC. The average UCs associated with Figs. \ref{fig:1a} and \ref{fig:1b} are $\overline{u_1}=(3\times 4+1\times 3)/4=3.75$ and $\overline{u_1}=(3\times 3+1\times 4)/4=3.25$, respectively.}\label{fig:1}
\end{figure}

Recently, there have been a lot of studies on LRCs. For example, in \cite{Wang_IntegerLRCBound_2015}, a tight upper bound on $d$ for LRCs with $\frac{n}{r} > \lceil\frac{n}{r+1}\rceil $ is proposed, and LRCs with the largest possible $d$ with $r\le \sqrt{n}-1$ are designed. 
In \cite{KamathREG_LRC2014}, LRCs with minimum storage regenerating and minimum bandwidth regenerating are considered. 
In \cite{goparaju_BLRC_2014,shahabi_BLRC15_10_2014,Constructions2016}, LRCs over small fields are proposed in order to decrease the computational complexity associated with coding.  

%In the case that $(r+1)\mid n$, among the total $n-k$ parity blocks corresponding with an optimal LRC, $\frac{n}{r+1}$ parity blocks are \textit{local} and the rest $(n-k-\frac{n}{r+1})$ parity blocks are \textit{global}. %\leq d-2-\floor{\frac{d-2}{r+1}}
%In the existing solutions, all the information blocks get involved in the generation of all global parity blocks  in order to satisfy the minimum distance constraint \cite{Silbe_optLRC_2013,barg_a_family_2014,OptLRCDimakis}. 
%Consequently, if some information blocks have to be updated, all the global parity blocks have to be changed resulting in a costly update process. 

For a systematic $(n,k,d,r)$ optimal LRC, there exist $n-k$ parity blocks constructed from $k$ information blocks. 
While some of these $n-k$ parity blocks are constructed locally from a few blocks to achieve the code locality, some other parity blocks are constructed globally to achieve the required minimum distance. In the existing optimal LRCs, all the information blocks get involved in these globally constructed parity blocks \cite{Silbe_optLRC_2013,barg_a_family_2014,OptLRCDimakis}.
Consequently, if some information blocks have to be updated, all the global parity blocks have to be changed resulting in a costly update process.

Is it possible to generate different optimal $(n,k,d,r)$ LRCs with different update complexity?
If yes, how can we find optimal LRCs with small update complexity? This is the central question studied in this paper.  

%To answer this question, we first need to define \textit{update complexity}. 
%In \cite{UC2010}, update complexity (UC) is defined as the maximum number of parity blocks needed to be changed when an information block is updated. 
%In the case that information blocks have to be updated frequently, high UC requires costly bandwidth.
%
%Reducing the number of information blocks involved in the parity blocks results in accessing and changing only \textit{some} parity blocks during an information block update. 	
Fig.~\ref{fig:1} shows the Tanner graphs of two optimal $(n,k,d,r)=(8,4,4,3)$ LRCs. 
In the LRC of Fig.~\ref{fig:1a}, all the information blocks are involved in the two parity blocks $g_1$ and $g_2$.  
However, in the LRC of Fig.~\ref{fig:1b}, only some information blocks are involved in $g_1$ and $g_2$.  
%Both Fig.~\ref{fig:1a} and Fig.~\ref{fig:1b} are an optimal $(n,k,d,r)=(8,4,4,3)$ LRC but with different structure. 
%Here, this structure change results an improvement in UC and therefore, for the same values of $d$, $r$, and code rate, LRC
In this example, if one information block is updated, the LRC of Fig.~\ref{fig:1b}, on average, needs $13\%$ less updates on the parity blocks. 

In this paper, we consider the problem of update complexity (UC) for systematic optimal LRCs. 
The contributions of this paper are two-folds. By taking an existing definition of update complexity and generalizing it, we obtain both the upper and lower bounds on UC for an importance class of LRCs.
Furthermore, we design a class of \textit{optimal} LRCs whose average UC is close to the obtained lower bound.
Note that this improvement of UC is achieved without sacrificing other important parameters of the LRC such as minimum distance ($d$), rate ($\frac{k}{n}$), or locality ($r$).

%Organization of the paper:
The remainder of this paper is organized as follows. In Section \ref{sec:preliminaries}, we provide the required preliminaries. In Section \ref{sec:BoundOnUpdateComplexity(UC)}, we obtain lower and upper bounds on UC. In Section \ref{sec:CodingMethodwithLowUC}, we introduce our proposed LRC with small UC and we compare our LRC with other LRCs in terms of UC. Finally, in Section \ref{sec:Conclusion}, we conclude the paper.   

\textbf{Notations:}
We show matrices and vectors by capital bold letters and bold letters, respectively.
$\mathbb{F}_q$ and $\otimes$ stand for a finite field of order $q$ and tensor product, respectively.
$\mathbf{I}_a$ and $\mathbf{0}_{b\times c}$ represent an identity matrix of size $a$ and a zero matrix of size $b\times c$, respectively.
$(\cdot)^T$ and $\mathbf{1}_a$ represent matrix transpose operation and a column vector of ones with size $a$, respectively. $\overline{u}$ represents $\frac{1}{n}\sum_{i=1}^{n} u_i$ which is the average of $u_i$'s for $i\in\{1,\cdots,n\}$. For an integer $n$, $[n]=\{1,\cdots,n\}$.

%% file: Sections/Preliminaries.tex
\subsection{Definitions}
\textit{Systematic linear block codes:} The generator matrix of an $(n,k)$ systematic linear block code can be presented as $\mathbf{G}=[\mathbf{I}_{k},\mathbf{P}]\in \mathbb{F}^{k\times n}_{q}$, where $\mathbf{P}\in\mathbb{F}^{k\times (n-k)}_{q}$. 
Assuming that $\mathbf{x}=[x_{1}, x_{2}, ..., x_{k}]\in\mathbb{F}^{1\times k}_{q}$ and $\mathbf{y}=[y_{1}, y_{2}, ..., y_{n}]\in\mathbb{F}^{1\times n}_{q}$ are the information and encoded vectors, respectively, we have $\mathbf{y}=\mathbf{x}\mathbf{G}$. 
The parity check matrix of the code is $\mathbf{H}=[-\mathbf{P}^{T},\mathbf{I}_{n-k}]\in \mathbb{F}^{(n-k)\times n}_{q}$ satisfying $\mathbf{G}\mathbf{H}^{T}=\mathbf{0}_{1\times(n-k)}$.   

\begin{remark}
In a DSS, in order to store a stripe of data of size $L$ symbols by an $(n,k)$ systematic linear block code, first, the stripe is partitioned into $k$ data blocks each of size $l=\frac{L}{k}$ symbols. 
Assume that $x_{i,j}$ is $i$-th symbol of $j$-th data block, where $i\in [l]$ and $j\in[k]$.
Then, $\mathbf{x}_i=[x_{i,1},\cdots,x_{i,k}]\in\mathbb{F}^{1\times k}_{q}$.
The coded vector $\mathbf{y}_i=[y_{i,1},\cdots,y_{i,n}]\in\mathbb{F}^{1\times n}_{q}$ is generated as $\mathbf{y}_i=\mathbf{x}_i\mathbf{G}=\mathbf{x}_i[\mathbf{I}_{k},\mathbf{P}]$. 
From which, matrix $\mathbf{Y}\in\mathbb{F}_q^{l\times n}$ is constructed by stacking $l$ encoded vectors $\mathbf{y}_i$.
Each column of $\mathbf{Y}$ is an encoded block which is stored in a data node.
For simplicity, from now on, we assume that $l=1$.
\end{remark}

\textit{Minimum distance of code $(d)$:} The minimum Hamming distance between any two codewords of an erasure code is defined as the minimum distance of that code, denoted $d$. 
Any $(n,k)$ erasure code with minimum distance $d$ tolerates any $d-1$ symbol erasures.

\textit{Locality:} 
For an $(n,k)$ linear block code, locality of the $i$-th encoded block where $i\in[n]$, denoted $r_i$, is defined as the minimum number of other blocks needed to participate in its recovery process. 
In other words, in the case that the $i$-th block is missing, at least $r_i$ other blocks are needed to reconstruct it. 
Locality of a code, denoted $r$, is defined as the maximum of $r_i$ for $i\in[n]$, i.e. $r=\max\limits_{i\in[n]}r_i$. 

\textit{Tanner/Factor graph:} A Tanner graph (Factor graph) \cite{Tannergraph1,Tannergraph2} associated with an $(n,k)$ linear block code is a bipartite graph with $n$ variable nodes on one side (usually shown by circles) and $n-k$ check nodes on the other side (usually shown by squares) which represent columns and rows of the code's parity check matrix $\mathbf{H}\in \mathbb{F}^{(n-k)\times n}_{q}$, respectively. 
There is an edge between $j$-th ($j\in[n]$) variable node and $i$-th ($i\in[n-k]$) check node in the Tanner graph, if $h_{i,j}$ is non-zero, where $h_{i,j}$ represent the element of $i$-th row and $j$-th column of $\mathbf{H}$. 

\textit{Information and parity nodes:}
Consider the Tanner graph associated with a systematic $(n,k)$ linear block code.
Among all the $n$ variable nodes, $k$ variable nodes are corresponding with $k$ information blocks.
We call these variable nodes \textit{information nodes} and represent them by white circles in the Tanner graph.
The remaining $n-k$  variable nodes are corresponding with $n-k$ parity blocks.
We call these variable nodes \textit{parity nodes} and represent them by shaded circles in the Tanner graph (see Fig. \ref{fig:1} as an example).

%\begin{figure}[t!]
%	\centering
%	\includegraphics[width=0.5\textwidth, scale=0.5]{genTanner1}
%	\caption{Construction of an optimal $(n,k,d,r)$ $\nu$-LRC. 
%		There are $\frac{n}{r+1}$ local and $d-2-\Big\lfloor\frac{d-2}{r+1}\Big\rfloor$ global check nodes, where $r+1$ is cardinality of each local group.}
%	\label{fig:nulrc}
%\end{figure}

\begin{figure*}[t!]
	\centering
	\includegraphics[width=0.95\textwidth, scale=0.6]{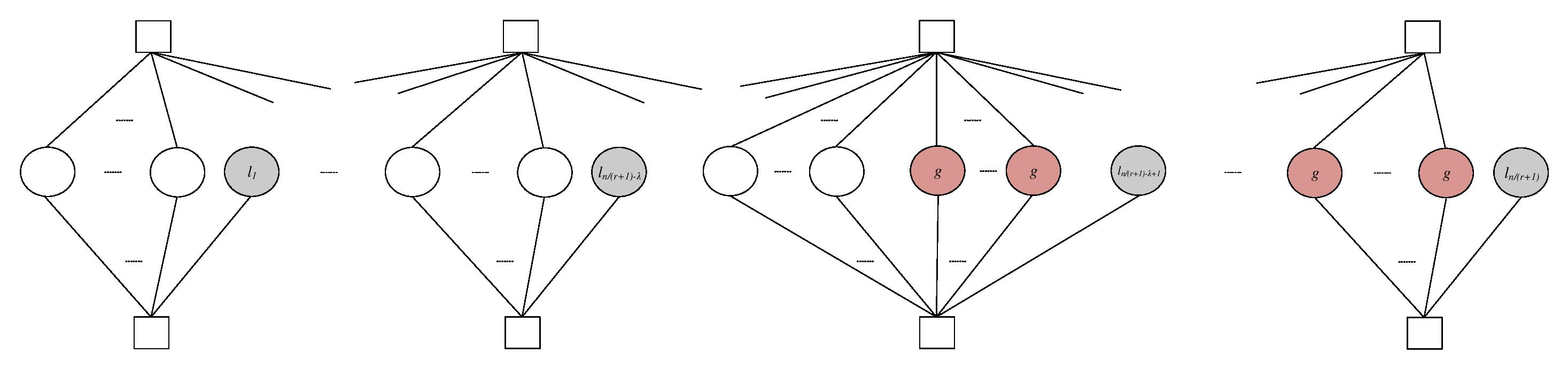}
	\caption{Construction of an $(n,k,d,r)$ $\nu$-LRC. 
		There are $\frac{n}{r+1}$ local and $d-2-\lfloor\frac{d-2}{r+1}\rfloor$ global parity nodes, where $r+1$ is cardinality of each local group.}
	\label{fig:genTanner}
\end{figure*}

\textit{Local and mixed groups, local and global check nodes and parity nodes:}
In the Tanner graph of an $(n,k,d,r)$ LRC, among $n-k$ check nodes, a minimal set of check nodes, each having at most $r+1$ edges that cover all variable nodes are called \textit{local check nodes} and constitute \textit{local groups}. 
In other words, a failed variable node can be reconstructed within its local group. Note that the locality $r_i$ of each variable is the size of its local group minus one.
Check nodes which are not local are called \textit{global check nodes}. 
Furthermore, parity nodes associated with local check nodes are called \textit{local parity nodes}; and the rest of parity nodes associated with global check nodes are called \textit{global parity nodes}. 
Local groups containing global parity nodes are called \textit{mixed groups}.
For example, in Figs. \ref{fig:1a} and \ref{fig:1b}, the check nodes that are below the variable nodes are local check nodes and the rest are global. 
In this example, the local group containing the fourth information node is a mixed group.

%Parity check nodes in LRC codes are also split to some local check nodes and some global check nodes. We assume that $r+1\mid n$, so in order to satisfy locality $r$ variable nodes are divided to ${\frac{n}{r+1}}$ groups and for each group a local parity CN is assigned and attached to all the group members. Furthermore, due to minimum distance $d$, global CNs which are $n-k-\frac{n}{r+1}$ nodes, should connected to all variable nodes. Each $(n,k)$ LRC has $k$ information nodes and also $n-k$ parity nodes. These $n-k$ parity blocks are distributed over the local groups and each group has one parity block to store the combination of information blocks. In the laso local group except the local parity blocks there are $d-2$ parity blocks which are corresponding to global parity blocks to store the combination of information blocks connected to global parity blocks. 
%}    

\subsection{An Important Class of Optimal LRCs}
\textit{Non-Overlapped and Uniform optimal LRCs ($\nu$-LRCs):} 
An important class of LRCs are optimal LRCs where $(r+1)\mid n$. 
The importance of this class stems from the fact that when $(r+1)\mid n$ the bound in (\ref{eq:minimumdistancebound}) can be achieved with equality. 
In fact, such optimal LRCs have been the focus of studies in \cite{OptLRCDimakis,Silbe_optLRC_2013,barg_a_family_2014,shahabi_2016_aclass}. 
Hence, we also assume $(r+1)\mid n$. 

Assuming that $(r+1)\mid n$, $\nu$-LRCs are a class of optimal LRCs in which $n$ encoded blocks are partitioned uniformly into $\frac{n}{r+1}$ non-overlapped local groups, where $r+1$ is the cardinality of each local group. 
In the structure of $\nu$-LRCs, local groups are both uniform and non-overlapped, hence the name.
In $\nu$-LRCs, among the total $(n-k)$ check nodes, there are $\frac{n}{r+1}$ local and $n-k-\frac{n}{r+1}$ global check nodes (see Fig. \ref{fig:genTanner} as an example).

In the following remark, the exact number of global check nodes for an $(n,k,d,r)$ $\nu$-LRC is computed.
\begin{remark}\label{remnulrc}
	Since $\nu$-LRC are optimal, by (\ref{eq:minimumdistancebound}), the number of their global check nodes can also be expressed as
\begin{equation*}
n-k-\frac{n}{r+1} = d-2-\Big\lfloor\frac{d-2}{r+1}\Big\rfloor.
\end{equation*}
\end{remark}

%According to Lemma~\ref{lem:parityLRC}, among the total number of $(n-k)$ parity block, there are $\frac{n}{r+1}=n-k-(d-2)$ local parity blocks. 
%As Fig.~\ref{fig:genTanner} shows, $n$ encoded blocks are partitioned into $\frac{n}{r+1}$ local groups, where $r+1$ is the cardinality of each local group.
%Each local group has a local parity block, denoted by $l_i$, where $i\in\{1,\cdots,\frac{n}{r+1}\}$.
%Also, except for the last local group, i.e. $\frac{n}{r+1}$-th local group, each local group has $r$ information blocks.
%The last local group has $(d-2)$ global parity blocks and one local parity block. 
%Thus, there are $r+1-(d-2)-1=r-d+2$ information blocks in the last local group.
%Observe that total number of information blocks is 
%\[
%k = (\frac{n}{r+1}-1)r+(r-d+2).
%\]

%% file: Sections/ProposedBound.tex
%In this section, upper and lower bounds on UC associated with optimal $(n,k,d,r)$ LRCs are established. Before presenting our results, we need to mention some lemmas and discuss some conditions required for the structure.
In this section, we formally define UC and then for the class of LRC that we discussed in the previous section, we find upper and lower bounds on UC.    

In \cite{UC2010}, update complexity (UC) is defined as the maximum number of parity blocks needed to be changed when an information block is updated. 
By generalizing this definition, we define UC, as the number of parity blocks needed to be changed when a set of $x$, $x\in[k]$, information blocks are updated, and denoted by $u_x$.

%\begin{figure}[t!]
%	\centering
%	\includegraphics[width=0.5\textwidth, scale=0.5]{rate}
%	\vspace{-.1 cm}
%	\caption{Relation between code rate and code size for the cases that $R> (1-\frac{1}{\sqrt{n}})^2$.}
%	\vspace{-.4 cm}
%	\label{fig:ratesize}
%\end{figure}

%An important class of LRCs are optimal LRCs where $(r+1)\mid n$. 
%The importance of this class stems from the fact that when $(r+1)\mid n$ the bound in (\ref{eq:minimumdistancebound}) can be achieved with equality. 
%In fact, such optimal LRCs have been the focus of study in \cite{OptLRCDimakis,Silbe_optLRC_2013,barg_a_family_2014,shahabi_2016_aclass}. 
%Hence, we also assume $(r+1)\mid n$
In this paper, we study the UC of $\nu$-LRCs and seek $\nu$-LRCs that have low UC.
In the following, we start with studying a special variable node arrangement for $\nu$-LRCs. Later, we will show how this arrangement helps us to establish our bounds.

Fig.~\ref{fig:genTanner} shows the general structure of an $(n,k,d,r)$ $\nu$-LRCs with a special variable node arrangement.
Here, $n$ encoded nodes are partitioned into $\frac{n}{r+1}$ local groups of size $r+1$.
Each local group has a local parity node, denoted by $l_i$, where $i\in[\frac{n}{r+1}]$.
Also, each local group except mixed groups has $r$ information nodes.
In general, the number of mixed groups can be arbitrarily large. However, in the structure shown in Fig. \ref{fig:genTanner}, the number of mixed group is minimal. This can lead to a small UC.
Also, there is at most one mixed group containing both information and global parity nodes.
We call this group \textit{infomixed group}\footnote{There is no infomixed group, if $r\mid k$.}.
As stated in Remark \ref{remnulrc}, there are $d-2-\lfloor\frac{d-2}{r+1}\rfloor$ global check nodes. 
Hence, the total number of mixed groups, denoted $\lambda$, is
\[
\lambda=\Big\lceil\frac{d-2-\lfloor\frac{d-2}{r+1}\rfloor}{r}\Big\rceil.
\]      
Consequently, the infomixed group, if exists, has $d-2-\lfloor\frac{d-2}{r+1}\rfloor-(\lambda-1)r$ global parity nodes and one local parity node. 
Thus, there are $r+1-(d-2-\lfloor\frac{d-2}{r+1}\rfloor-(\lambda-1)r)-1$ information nodes in the infomixed group.
Observe that the total number of information nodes is 
\[
k = (\frac{n}{r+1}-\lambda)r+(r-d+2+\Big\lfloor\frac{d-2}{r+1}\Big\rfloor+(\lambda-1)r).
\]
where the first term is the number of information nodes in none-mixed groups and the second term is that in infomixed group.
Now by using the given properties of $\nu$-LRCs, in the following theorem, we establish both lower and upper bounds on UC.

\begin{thrm}\label{thrm:UCbound}
	For an $(n,k,d,r)$ $\nu$-LRC, UC associated with changing $x$ information blocks ($u_x$) is bounded as 
	\[d-1+ \Big\lceil\frac{x-(r\lambda-(d-2-\lfloor\frac{d-2}{r+1}\rfloor))}{r}\Big\rceil \leq u_{x} \le d-1+\theta,\]
	where $\lambda=\Big\lceil\frac{d-2-\lfloor\frac{d-2}{r+1}\rfloor}{r}\Big\rceil$, and 
	\[
	\theta =
	\begin{cases}
	x & \text{if } x\leq \frac{n}{r+1}-\lambda\\
	\frac{n}{r+1}-\lambda & \text{otherwise}
	\end{cases}.
	\]
\end{thrm}
\begin{IEEEproof}   
	Please refer to Appendix \ref{app:Proof_UCbound}.
	%Please see the arXive version of paper in \cite{Proofs}.
\end{IEEEproof}

%Comparing different codes based on the UC of each information block is not a efficient way to measure the update complexity of each code.
%Hence, we use average UC to compare different codes in terms of their update complexity. 
Update complexity associated with a set of $x$ information blocks ($u_x$), can vary from one set to another set, both with cardinality $x$.
Hence, another measure of the update complexity of a code is needed.
We use the average UC as a metric to evaluate UC of a code.
%The average UC of code is another measure of the update complexity to measure the update complexity of information blocks.
Average UC of code, denoted $\overline{u_x}$, is defined as the average number of parity blocks needed to be changed when any set of $x$ information blocks, where $x \in [k]$, are updated. 
In this work, our focus is on $\overline{u_1}$, which is the average update complexity when only one information block is changed.
Observe that $\overline{u_x}$ can be enhanced by improving $\overline{u_1}$. 
In the following theorem, both lower and upper bounds on $\overline{u_1}$ is computed.
%In order to find upper and lower bounds on average UC, we use the established bounds in Theorem \ref{thrm:UCbound} and by modifying and making that bounds tighter for the case that only one information block is updated, we establish upper and lower bound on $\overline{u_1}$. 
%In this case from Theorem \ref{thrm:UCbound}, we have

%In order to establish a tight bound on $\overline{u_1}$, we need to find the minimum possible number of information nodes connected to $d$ parity nodes in the structure of $\nu$-LRCs.

%Later in the next section, we propose a class of optimal LRCs that has the minimum number of information nodes with update complexity $d$.
%To find the minimum possible number of information nodes connected to $d$ parity nodes, we use the following theorem from \cite{shahabi_ravg_2016} to ensure that the minimum distance constraint is satisfied.

%In the following remark, the necessary and sufficient conditions on $d$ is stated. 

\begin{thrm}\label{thrm:aveUC1}
	For an $(n,k,d,r)$ $\nu$-LRC, average UC associated with changing one information block ($\overline{u_1}$) is bounded as
	\[(d-1)+\frac{\eta\Big\lfloor\frac{\frac{n}{r+1}-\lambda}{\lfloor\frac{d-2}{r+1}\rfloor+1}\Big\rfloor}{k}\le \overline{u_1} \le d\]	
where $\lambda=\Big\lceil\frac{d-2-\lfloor\frac{d-2}{r+1}\rfloor}{r}\Big\rceil$, 
	\[
	\eta =
	\begin{cases}
	0 & \text{if } \frac{\alpha}{(d-2-\lfloor\frac{d-2}{r+1}\rfloor)}\mid \beta ~\text{or}~ \lfloor\frac{d-2}{r+1}\rfloor=\lceil\frac{d-2}{r+1}\rceil\\
	\alpha-\beta\lfloor\frac{\alpha}{\beta}\rfloor & \text{otherwise}
	\end{cases},
	\]
$\alpha=((r+1)(\lfloor\frac{d-2}{r+1}\rfloor+1)-(d-2))(d-2-\lfloor\frac{d-2}{r+1}\rfloor)$, and $\beta=r(\lfloor\frac{d-2}{r+1}\rfloor+1)$.
\end{thrm}
\begin{IEEEproof}
	Please refer to Appendix \ref{app:Proof_UC1}.
	%Please see the arXive version of paper in \cite{Proofs}.
\end{IEEEproof}
%Thus, for the total number of these information node, denoted $z_{total}$, we have
%\begin{equation}\label{modifiedbound}
%\min z_{total} = z\Big\lfloor\frac{\frac{n}{r+1}-\lambda}{\lfloor\frac{d-2}{r+1}\rfloor+1}\Big\rfloor
%\end{equation}
%Now, by considering (\ref{modifiedbound}) for the case that there is one information node update, bounds on average UC ($\overline{u_1}$) can be established as
%\begin{equation*}\label{modifiedbounduc}
%%~~~~~~~\frac{z_{total}d+(k-z_{total})(d-1)}{k}\le \overline{u_1} \le \frac{kd}{k} \Rightarrow \\
%(d-1)+\frac{z_{total}}{k}\le \overline{u_1} \le d
%\end{equation*}

In the following section, we propose a class of $\nu$-LRCs with small average UC, close or even in some cases equal to the lower bound on $\overline{u_1}$.

%% file: Sections/ProposedLRC.tex
Here, we present our proposed LRCs using their Tanner graphs. 
Our proposed LRCs achieve the bound in (\ref{eq:minimumdistancebound}), i.e. they are optimal. 
Furthermore, they benefit from a small $\overline{u_1}$, close or even equal to the lower bound obtained in Theorem \ref{thrm:aveUC1}.
In other words, in comparison with the existing optimal LRCs, our proposed LRCs require accessing and changing a smaller number of parity blocks in the case of information block updates.  
%Hence, these two important properties of our proposed LRCs, i.e. optimality and small UC, are of interest from a practical point of view considering the application of LRCs in real-world DSSs. 

%In order to achieve minimum distance $d$, the necessary and sufficient conditions on $d$ in the following remark should be satisfied.

\subsection{Construction of Our Proposed Optimal LRCs}
In order to construct our proposed LRCs, first, $n$ variable nodes are partitioned into $\frac{n}{r+1}$ local groups each containing $r+1$ variable nodes. 
Hence, there are $\frac{n}{r+1}$ local check nodes associated with $\frac{n}{r+1}$ local groups, where each local group constructs one local parity block.
The remaining $n-k-(\frac{n}{r+1})=d-2-\lfloor\frac{d-2}{r+1}\rfloor$ check nodes construct $d-2-\lfloor\frac{d-2}{r+1}\rfloor$ global parity blocks which are placed in the mixed groups (Fig. \ref{fig:genTanner}).

By Theorem \ref{thrm:UCbound}, we have 
\begin{equation*}
d-1\le u_1 \le d.
\end{equation*} 
This implies that updating an information block requires updating either $d-1$ or $d$ parity blocks.
Therefore, to minimize $\overline{u_1}$, we have to find tanner graphs with minimum number of information nodes whose update requires changing $d$ parity nodes.
While constructing such tanner graphs, we have to ensure the minimum distance constraint is satisfied, and for that we use the following theorem from \cite{shahabi_ravg_2016}.
\begin{thrm}\label{thm:con_d}
	\cite{shahabi_ravg_2016} There is an erasure code with minimum distance $d$ associated to Tanner graph $\mathcal{T}$ iff every $\gamma$ check nodes of $\mathcal{T}$ cover $\gamma+k$ variable nodes, where $\gamma\in[n-k-d+2,n-k]$.
\end{thrm}

By Theorem \ref{thm:con_d}, a necessary condition to guarantee the minimum distance $d$ for our proposed LRCs is that any collection of $n-k-d+2$ check nodes consisting of $n-k-d+1$ local check nodes and a single global check node must cover at least $n-(d-2)$ variable nodes. 
The number of local groups outside the selected collection is $\frac{n}{r+1}-(n-k-d+1)=\lfloor\frac{d-2}{r+1}\rfloor+1$. 
To satisfy this condition, the single global check node in any such collections must be connected to all the variable nodes of the local groups outside the collection with at most $d-2$ exceptions. 
In other words, at least $(r+1)(\lfloor\frac{d-2}{r+1}\rfloor+1)-(d-2)$ variable nodes of any set of $\lfloor\frac{d-2}{r+1}\rfloor+1$ local groups have to be connected to each of the $d-2-\lfloor\frac{d-2}{r+1}\rfloor$ global check nodes.

Considering this, our proposed LRCs can be constructed using Algorithm \ref{alg} presented in the next page. 
Note that by choosing coefficients of parity check matrix associated  with the obtained Tanner graph randomly from a sufficiently large Galois field, the LRC can be generated.

%Now, by considering these statements, our proposed optimal LRCs are constructed by Algorithm \ref{alg}.
%By following Algorithm \ref{alg}, we are able to construct a structure with improved UC and based on Theorem \ref{thm:con_d} there exists an erasure code for the obtained structure. 
%Hence, by choosing the coefficients of the information blocks of the obtained structure from a sufficient large Galois field we can generate our code.    

\begin{algorithm}[t]
	First, construct local groups based on the structure of a $\nu$-LRC depicted in Fig.~\ref{fig:genTanner} \textbf{then} \\
	$\bullet$ Connect each of the $d-2-\lfloor\frac{d-2}{r+1}\rfloor$ global check node to a distinct global parity node located in the mixed groups. \\
	$\bullet$ Connect all the information nodes of the infomixed group, if exists, to all the $d-2-\lfloor\frac{d-2}{r+1}\rfloor$ global check nodes. \textbf{then} \\
	\For{$i\in[d-2-\lfloor\frac{d-2}{r+1}\rfloor]$,}
	{\While{There is a set of $\lfloor\frac{d-2}{r+1}\rfloor+1$ local groups out of all $\frac{n}{r+1}$ local groups, where $i$-th global check node is not connected to at least $(r+1)(\lfloor\frac{d-2}{r+1}\rfloor+1)-(d-2)$ variable nodes of that set}{Connect the $i$-th global check node to $(r+1)(\lfloor\frac{d-2}{r+1}\rfloor+1)-(d-2)$ information nodes of the selected set.}}
	
%	\While{There is $i\in[d-2-\lfloor\frac{d-2}{r+1}\rfloor]$ and $j_1,\dots,j_{\lfloor\frac{d-2}{r+1}\rfloor+1}\in[\frac{n}{r+1}]$ where $i$-th global check node is not connected to at least $(r+1)(\lfloor\frac{d-2}{r+1}\rfloor+1)-(d-2)$ variable nodes of $j_1$-th$-j_{\lfloor\frac{d-2}{r+1}\rfloor+1}$-th local groups}
%	%	{\eIf{$j_m < (\frac{n}{r+1}-\lambda)$ where $m\in\{1,\dots,\lfloor\frac{d-2}{r+1}\rfloor+1\}$}
%	%		{$\bullet$ connect each of the $d-2$ global parity nodes in $j$-th local group to $i$-th global check node\;
%	%			
%	%			$\bullet$ connect all information nodes in $j$-th local group to $i$-th global check node\;}		
%	%	}
%	{connect $i$-th global check node to $(r+1)(\lfloor\frac{d-2}{r+1}\rfloor+1)-(d-2)$ information nodes of $j_1$-th$-j_{\lfloor\frac{d-2}{r+1}\rfloor+1}$-th local groups except the information nodes of infomixed group which already have required connections with global check nodes.}
	\If{There is an information node in the non-mixed groups connected to $\beta$ global check nodes, where $\beta<(d-2-\lfloor\frac{d-2}{r+1}\rfloor-(\lambda-\lfloor\frac{d-2}{r+1}\rfloor))$ and $\lambda=\Big\lceil\frac{d-2-\lfloor\frac{d-2}{r+1}\rfloor}{r}\Big\rceil$}{connect it to $d-2-\lambda-\beta$ global check nodes in order to have connection with at least $d-2-\lfloor\frac{d-2}{r+1}\rfloor-(\lambda-\lfloor\frac{d-2}{r+1}\rfloor)$ global check nodes}
	\caption{Construction of optimal LRCs with small UC}
	\label{alg}
\end{algorithm}

\subsection{Properties and Evaluation of Our Proposed Optimal LRCs}
In the following, we verify some important properties of our proposed LRCs generated by Algorithm \ref{alg}.

%\begin{enumerate}
%	\item minimum distance constraint, and 
%	\item small UC close to lower bound in~\ref{prop:UCbound}.
%\end{enumerate}

\begin{prop}\label{prop:min_d}
The proposed optimal $(n,k,d,r)$ LRCs constructed by Algorithm \ref{alg} have minimum distance $d$.
\end{prop}
\begin{IEEEproof}
%Please see the arXive version of paper in \cite{Proofs}.
Please refer to Appendix \ref{app:Proof_d}.
\end{IEEEproof}

\begin{remark}
Our proposed LRC is optimal since all the assumptions in the construction of 
our proposed LRCs are made based on the satisfaction of bound in 
(\ref{eq:minimumdistancebound}).
\end{remark}

%	Considering Lemma \ref{lem:parityLRC}, our proposed LRC achieves bound in 
%(\ref{eq:minimumdistancebound}) when $(r+1)\mid n$ and $(d-2)<(r+1)$.   
%	In other words, our proposed $(n,k,d,r)$ LRC is optimal.
%the number of local and global check nodes in our proposed LRC is chosen based 
%on Lemma \ref{lem:parityLRC}, in which it is assumed that, bound in 
%\ref{eq:minimumdistancebound} is achieved. 

\begin{remark}
%	In the Tanner graph of our proposed LRC, \textit{not} all t he information nodes are connected to the global check nodes. 
%	Therefore, maximum possible number of information nodes in the Tanner graph are associated with $d-1$ parity nodes and other information nodes are associated with $d$ parity.    
%	Hence, due to generating more information nodes, which are associated with $d-1$ parity nodes, than other LRCs method, our proposed $(n,k,d,r)$ LRC has small UC close to the lower bound of Proposition \ref{prop:UCbound} 
Our proposed optimal LRCs improve the UC compared to the existing solutions. 	
The construction of our proposed LRCs ensures that not all information nodes be involved in global check nodes. 
In fact, it tries to keep the number of information nodes involved in any given global check node small. 
This means a small number of global parity blocks need update when information blocks are updated. 
This in turn results in codes with average UC close to the lower bound.  
\end{remark}

\begin{remark}\label{rem:balance}
In the case that all information nodes in the non-mixed groups have the same degree and each of them is connected to exactly $d-2-\lfloor\frac{d-2}{r+1}\rfloor-(\lambda-\lfloor\frac{d-2}{r+1}\rfloor)$ global check nodes, the lower bound of single updates ($u_1=d-1$) for all information nodes is achieved.
For our proposed optimal $(n,k,d,r)$ LRC, this is the case when $((r+1)(\lfloor\frac{d-2}{r+1}\rfloor+1)-(d-2))\mid \beta$, where $\beta=r(\lfloor\frac{d-2}{r+1}\rfloor+1)$, is satisfied.
\end{remark}

%\begin{remark} \label{remarkblance}
%In the case that all information nodes in the first $\frac{n}{r+1}-\lambda$ local groups have the same degree and each of them is connected to exactly $d-3-\lfloor\frac{d-2}{r+1}\rfloor$ global check nodes, lower bound of single update for all information nodes is achieved. 
%For our proposed optimal $(n,k,d,r)$ LRC, this is the case when $(d-3)\mid r$ is satisfied.  
%This is because, if in our proposed LRC
%	\[(r+1)-(d-2)=(r-d+3)\mid r \]
%is satisfied, then in order to connect each global check node to a local group, it is possible to find exactly $r-d+3$ information nodes out of $r$ information nodes with the lowest degree. 
%In other words, $(r-d+3)+1$ information nodes do not have the same degree except when all the information nodes in a local group are not connected to all global check nodes. 
%Thus, all the information nodes in the first $n-k-d+1$ local groups have degree $d-3+1=d-2$.    	
%\end{remark}

\begin{figure}[t!]
	\centering
	\includegraphics[width=0.5\textwidth, scale=0.5]{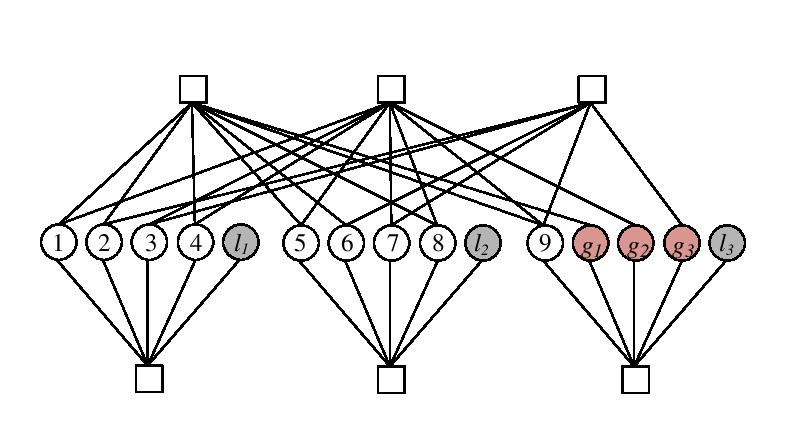}
	\vspace{-.5 cm}
	\caption{Tanner graph of an $(n,k,d,r)=(15,9,5,4)$ optimal LRC. In this figure, the global check nodes are connected to variable nodes based on our proposed method.}
	\vspace{-.4 cm}
	\label{fig:2}
\end{figure}

%In the Fig.\ref{fig:2}, we construct an optimal $(n,k,d,r)=(15,9,5,4)$ LRC using Algorithm \ref{alg}.
\begin{example}\label{ex:ex}
	Tanner graph of an optimal $(n,k,d,r)=(15,9,5,4)$ LRC is obtained using the Algorithm \ref{alg} (see Fig. \ref{fig:2}). In this example, we have 
	\[((r+1)(\lfloor\frac{d-2}{r+1}\rfloor+1)-(d-2))=2\mid \beta=4.\]	
	Hence, Remark \ref{rem:balance} is satisfied and all the information nodes are connected to $d-1=4$ parity nodes. 
\end{example}

\begin{figure}[t!]
	\centering
	\includegraphics[width=0.5\textwidth, scale=0.5]{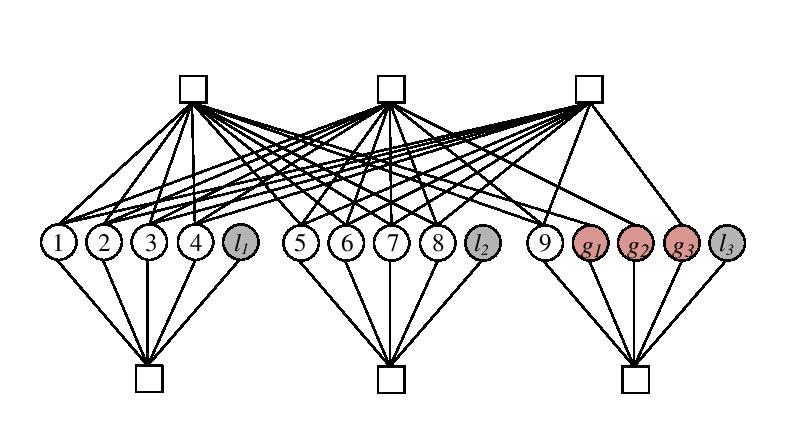}
	\vspace{-.5 cm}
	\caption{Tanner graph of an $(n,k,d,r)=(15,9,5,4)$ optimal LRC. In this figure, each of the global check node are connected to $k+1=10$ variable nodes.}
	\vspace{-.4 cm}
	\label{fig:3}
\end{figure}

In order to evaluate UC of our proposed coding scheme, we compare our proposed $(n,k,d,r)=(15,9,5,4)$ optimal LRC presented in Example \ref{ex:ex} with one whose global check nodes are connected to $k+1$ variable nodes, which is the common approach in designing optimal LRCs (please see Fig. \ref{fig:3}).


The average UC associated with each LRC is computed when $1$ and $2$ information nodes update. 
In the following equations, we denote the average UC of our proposed LRC and that of other LRC by $\overline{u_x}^{LRC_1}$ and $\overline{u_x}^{LRC_2}$, respectively.
For $u_1$, we have 
\[ 4\le {u_{1}}\le 5,\]
\[\overline{u_{1}}^{LRC1}=\frac{9(d-1)}{9}=d-1=4\text{, and}\]
\[\overline{u_{1}}^{LRC2}=\frac{(d-1)+8d}{9}=d-0.11=4.89.\]
Similarly, for $u_2$, we have 
\[5\le {u_{2}}\le 6,\]
\[\overline{u_{2}}^{LRC1}=(26\times 5 + 10 \times 6)/{9\choose 2}=5.27\text{, and}\]
\[\overline{u_{2}}^{LRC2}=(20\times 5 + 16 \times 6)/{9\choose 2}=5.44.\]
 
Hence, our proposed optimal LRC suggests UC close to the lower bound obtained in Theorem \ref{thrm:UCbound}. 
In this case, our proposed optimal LRC improves the UC by 18.2\% and 3.1\% for $\overline{u_1}$ and $\overline{u_2}$, respectively. 
Comparing with other approaches of designing optimal LRCs, this improvement is obtained without sacrificing other important properties of LRCs.

\begin{figure}[t!]
	\centering
	\includegraphics[width=0.5\textwidth, scale=0.5]{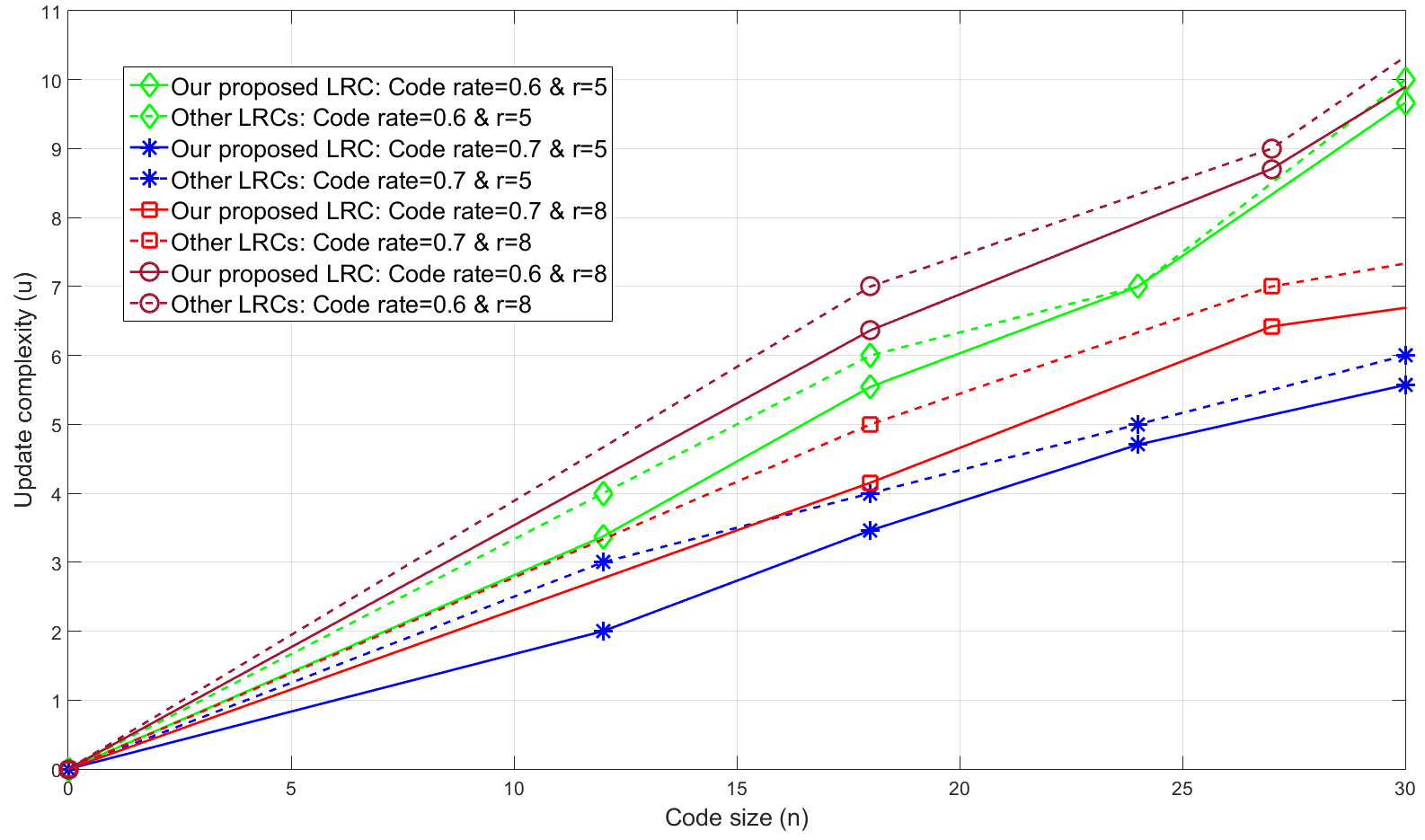}
	\vspace{-.5 cm}
	\caption{Comparison between the average UC of our proposed LRCs and other LRCs for two different code rate and code locality.}
	\vspace{-.4 cm}
	\label{fig:4}
\end{figure}

In Fig. \ref{fig:4}, we compare the general approach of LRCs and our proposed optimal LRCs in terms of two different code rates and code localities.

%% file: Sections/Conclusion.tex
The class of locally repairable codes (LRCs) is an important class of erasure codes to store data efficiently in distributed storage systems.
In this paper, we established bounds on the update complexity (UC) of an important class of LRCs.
Furthermore, we proposed a class of LRCs with small UC.
Considering the recent usage of LRCs in practice, e.g. in Facebook HDFS-RAID and Windows Azure Storage, our proposed LRCs could also be of interest from a practical point of view.

%% file: Sections/boundUC.tex
Here, we prove Theorem \ref{thrm:UCbound}. 
First, we state a lemma to find the minimum parity nodes required to be 
connected to each information node.

\begin{lem}\label{lem:gen}
	In the Tanner graph associated with an $(n,k,d,r)$ $\nu$-LRC shown 
	in Fig.~\ref{fig:genTanner}, any information node of non-infomixed and 
	infomixed groups is linked to at least $d-1$ and exactly $d-1$ parity 
	nodes, respectively.
\end{lem}
\begin{IEEEproof}
	The generator matrix associated with the $\nu$-LRC related to Fig.~\ref{fig:genTanner} is
	\begin{equation*}			
	\mathbf{G}=[\mathbf{I}_k, \mathbf{P}]\in \mathbb{F}^{k\times n}_{q},
	\end{equation*}
	where $\mathbf{P} = [\mathbf{P}_1, \mathbf{P}_2]\in 
	\mathbb{F}^{k\times(n-k)}_{q}$ is the parity matrix generator. 
	Matrix $\mathbf{P}_1\in\mathbb{F}^{k\times (\frac{n}{r+1}-\lambda)}_{q}$, 
	which generates the local parity nodes of the first $\frac{n}{r+1}-\lambda$ 
	local groups, can be presented as 
	\begin{equation*}
	\mathbf{P}_1= \left[\begin{array}{cc}
	\mathbf{I}_{\frac{n}{r+1}-\lambda}\otimes\mathbf{1}_{r}\\
	\mathbf{0}_{(r-d+2+\lfloor\frac{d-2}{r+1}\rfloor+(\lambda-1)r)\times(\frac{n}{r+1}-\lambda)}
	\end{array}\right]\in\mathbb{F}^{k\times (\frac{n}{r+1}-\lambda)}_{q}.
	\end{equation*}
	As well, matrix $\mathbf{P}_2\in\mathbb{F}^{k\times 
	(d-2-\lfloor\frac{d-2}{r+1}\rfloor+\lambda)}_q$ generates the local parity 
	nodes of the $\lambda$ mixed groups, and the global parity nodes. 
	Observe that the $i$-th row of $\mathbf{P}_1$ has one non-zero element for 
	$i\in[k-(r\lambda-d+2+\lfloor\frac{d-2}{r+1}\rfloor)]$ and the last 
	$r\lambda-d+2+\lfloor\frac{d-2}{r+1}\rfloor$ rows of $\mathbf{P}_1$ are 
	all zero.
	In order to satisfy the minimum distance constraint, each row of 
	$\mathbf{G}$ must have at least $d$ non-zero elements. 
	Considering the first $k-(r\lambda-d+2+\lfloor\frac{d-2}{r+1}\rfloor)$ rows of $\mathbf{G}$, each row has two nonzero elements from $\mathbf{I}_k$ and $\mathbf{P}_1$. 
	Thus, each row of $\mathbf{P}_2$ must have at least $d-2$ non-zero elements. 
	In other words, each row of $\mathbf{P}_2$ has at most one zero element. 
	Note that the last $r\lambda-d+2+\lfloor\frac{d-2}{r+1}\rfloor$ rows of $\mathbf{P}_1$ have no non-zero elements. 
	Hence, at least $d-1$ elements of the last $r\lambda-d+2+\lfloor\frac{d-2}{r+1}\rfloor$ rows of $\mathbf{P}_2$ are non-zero.    
	Note that if $r\lambda-d+2+\lfloor\frac{d-2}{r+1}\rfloor>0$,  i.e. there 
	exists an infomixed group, we have 
	$d-2-\lfloor\frac{d-2}{r+1}\rfloor+\lambda=d-1$.
	Thus, $\mathbf{P}_2$ has 
	$d-1$ columns; and all $d-1$ elements of the last 
	$r\lambda-d+2+\lfloor\frac{d-2}{r+1}\rfloor$ rows of $\mathbf{P}_2$ are 
	non-zero. 
\end{IEEEproof}

\vspace{2mm}
\textit{Proof of Lower bound:}   

According to Lemma \ref{lem:gen}, each row of $\mathbf{P}$ has at least $d-1$ 
non-zero elements.
Thus, any information node update leads to at least $d-1$ parity node updates. 
Regarding Fig. \ref{fig:genTanner}, the last 
$r\lambda-d+2+\lfloor\frac{d-2}{r+1}\rfloor$ information nodes are involved 
in the $\lambda$ local parity nodes and $d-2-\lfloor\frac{d-2}{r+1}\rfloor$ 
global parity nodes.
Hence, information nodes associated with the infomixed group require exactly 
$d-1$ parity node updates which is the minimum required updates. 
Now, assume that the number of information nodes to be updated is less than or 
equal to the number of information nodes associated with the infomixed group, 
i.e $x\leq r\lambda-d+2+\lfloor\frac{d-2}{r+1}\rfloor$.
Then, assuming all the $x$ information nodes are in the infomixed group, the 
lower bound on UC associated with $x$ information node updates is $(d-1)$. 
On the other hand, if the number of information nodes to be updated exceeds the 
number of information nodes associated with the infomixed group, i.e if 
$x>r\lambda-d+2+\lfloor\frac{d-2}{r+1}\rfloor$, 
we assume that $r\lambda-d+2+\lfloor\frac{d-2}{r+1}\rfloor$ information nodes 
are in the infomixed group and the remaining 
$x-r\lambda-d+2+\lfloor\frac{d-2}{r+1}\rfloor$ information nodes are in other 
local groups. 
Note that if an information node associated with the first 
$\frac{n}{r+1}-\lambda$ local groups is updated, then its local parity node 
has to be updated too.
Thus, in this case, the total number of local parity nodes which have to be 
updated is 
$\Big\lceil\frac{x-(r\lambda-d+2+\lfloor\frac{d-2}{r+1}\rfloor)}{r}\Big\rceil$ and the 
lower bound on $u_x$ is obtained as
$u_{x} \ge d-1+ \Big\lceil\frac{x-(r\lambda-d+2+\lfloor\frac{d-2}{r+1}\rfloor)}{r}\Big\rceil$.

\vspace{2mm}
\textit{Proof of Upper bound:}   

In order to obtain the upper bound on $u_x$, we assume that $x$ information 
nodes are located in distinct $x$ local groups. 
From Lemma \ref{lem:gen}, any information node update requires at least $d-1$ 
parity node updates. 
Assuming that $x\le \frac{n}{r+1}-\lambda$, among all $x$ information node 
updates, there exists only one information node update in each local group. 
Thus, in this case, the number of local parity node updates is exactly $x$. 
On the other hand, if $x>\frac{n}{r+1}-\lambda$, each of the first 
$\frac{n}{r+1}-\lambda$ local groups has exactly one local parity node update. 
Hence, the upper bound on $u_x$ is obtained as $u_{x} \le d-1+\theta$, 
where $\theta=x$ if $x\le \frac{n}{r+1}-\lambda$ and $\theta=\frac{n}{r+1}-\lambda$ otherwise.
Observe that for $x>\frac{n}{r+1}-\lambda$, $d-1+\theta=n-k$ which is the amount of all parity nodes.

%% file: Sections/Proof_UC1.tex
Here, we prove Theorem \ref{thrm:aveUC1}.
By Theorem \ref{thrm:UCbound}, we know that updating an information node 
requires updating either $d-1$ or $d$ parity nodes.
In the following, first, we state a lemma to find the minimum number of 
information nodes connected to $d$ parity nodes in any set of 
$\lfloor\frac{d-2}{r+1}\rfloor+1$ non-mixed groups.
Then, by using this lemma, we prove the established lower and upper bounds.

\begin{lem}\label{lem:min_d}
Within any collection of $\lfloor\frac{d-2}{r+1}\rfloor+1$ local groups in an $(n,k,d,r)$ $\nu$-LRC shown in Fig.~\ref{fig:genTanner}, the minimum number of information nodes except the information nodes of the infomixed group, denoted $\eta$, with UC equal to exactly $d$ is
	\[
	\eta =
	\begin{cases}
	0 & \text{if } \frac{\alpha}{(d-2-\lfloor\frac{d-2}{r+1}\rfloor)}\mid \beta ~\text{or}~ \lfloor\frac{d-2}{r+1}\rfloor=\lceil\frac{d-2}{r+1}\rceil\\
	\alpha-\beta\lfloor\frac{\alpha}{\beta}\rfloor & \text{otherwise}
	\end{cases},
	\]
	where 
	$\alpha=((r+1)(\lfloor\frac{d-2}{r+1}\rfloor+1)-(d-2))(d-2-\lfloor\frac{d-2}{r+1}\rfloor)$
	 and $\beta=r(\lfloor\frac{d-2}{r+1}\rfloor+1)$.
\end{lem}
\begin{IEEEproof}
	In an $(n,k,d,r)$ $\nu$-LRC, there are 
	$d-2-\lfloor\frac{d-2}{r+1}\rfloor$ global check nodes. 
	By Theorem \ref{thm:con_d}, a necessary condition to guarantee a minimum 
	distance $d$ for our proposed codes is that any collection of check nodes 
	consisting of $n-k-d+1$ 
	local check nodes and a single global check node must cover at least 
	$n-(d-2)$ variable nodes.
	Hence, each of the global check nodes have to be connected to at least $(r+1)(\lfloor\frac{d-2}{r+1}\rfloor+1)-(d-2)$ information nodes in any set of $\lfloor\frac{d-2}{r+1}\rfloor+1$ local groups. 
	Consequently, for every set of $\lfloor\frac{d-2}{r+1}\rfloor+1$ local 
	groups, there exist at least 
	$\alpha=((r+1)(\lfloor\frac{d-2}{r+1}\rfloor+1)-(d-2))(d-2-\lfloor\frac{d-2}{r+1}\rfloor)$
	 connections between the information nodes and the global check nodes. 
	Also, there exist $r$ information nodes in each local group but the mixed 
	groups.
	Thus, in a set of $\lfloor\frac{d-2}{r+1}\rfloor+1$ non-mixed groups, we 
	have $\beta=r(\lfloor\frac{d-2}{r+1}\rfloor+1)$ information nodes.  
	Consequently, in a set of $\lfloor\frac{d-2}{r+1}\rfloor+1$ non-mixed 
	groups, each information node has to be connected to at least 
	$\lfloor\frac{\alpha}{\beta}\rfloor$ global check nodes;
	hence, for all the information nodes in this set, we need at least 
	$\beta\lfloor\frac{\alpha}{\beta}\rfloor$ connections to global check 
	nodes. 
	Therefore, by subtracting $\beta\lfloor\frac{\alpha}{\beta}\rfloor$ from 
	all the $\alpha$ connections, we obtain the minimum number of information 
	nodes connected to $d$ parity nodes ($\eta$) in a set of 
	$\lfloor\frac{d-2}{r+1}\rfloor+1$ non-mixed groups as
	\begin{equation*}
	\eta=\alpha-\beta\Big\lfloor\frac{\alpha}{\beta}\Big\rfloor.
	\end{equation*}
	
	Also, if $\lfloor\frac{d-2}{r+1}\rfloor=\lceil\frac{d-2}{r+1}\rceil$, there 
	is no infomixed group; and all the information nodes are connected to 
	exactly $d-1$ parity nodes, and consequently, $\eta=0$.
	Furthermore, by Remark \ref{rem:balance}, if 
	$\frac{\alpha}{(d-2-\lfloor\frac{d-2}{r+1}\rfloor)}=((r+1)(\lfloor\frac{d-2}{r+1}\rfloor+1)-(d-2))\mid
	 \beta$, the same as the previous case, all the information nodes are 
	 connected to exactly $d-1$ parity nodes, and consequently, $\eta=0$. 
\end{IEEEproof}
By Lemma \ref{lem:min_d}, the minimum total number of information nodes 
connected to exactly $d$ parity nodes is equal to 
$\eta\Big\lfloor\frac{\frac{n}{r+1}-\lambda}{\lfloor\frac{d-2}{r+1}\rfloor+1}\Big\rfloor$.
Hence, we have
\[\frac{\eta\Big\lfloor\frac{\frac{n}{r+1}-\lambda}{\lfloor\frac{d-2}{r+1}\rfloor+1}\Big\rfloor
d+(k-\eta\Big\lfloor\frac{\frac{n}{r+1}-\lambda}{\lfloor\frac{d-2}{r+1}\rfloor+1}\Big\rfloor)(d-1)}{k}\le
 \overline{u_1} \le \frac{kd}{k}.\]
Therefore, 
\[(d-1)+\frac{\eta\Big\lfloor\frac{\frac{n}{r+1}-\lambda}{\lfloor\frac{d-2}{r+1}\rfloor+1}\Big\rfloor}{k}\le
 \overline{u_1} \le d.\]

%% file: Sections/proof_d.tex
Here, we prove Proposition \ref{prop:min_d}.\balance
For an $(n,k)$ erasure code with Tanner graph $\mathcal{T}$, if any $\varphi$ variable nodes are connected to $\varphi$ distinct check nodes, where $\varphi \in [d-1]$, then any $\varphi$ variable nodes can be recovered using equations associated with the distinct check nodes.
This implies that the code can recover up to any $d-1$ failures and therefore, it has minimum distance $d$.

In our proposed LRCs, every information nodes is connected to at least $(d-2-\lfloor\frac{d-2}{r+1}\rfloor-(\lambda-\lfloor\frac{d-2}{r+1}\rfloor)+\lambda)+1=d-1$ distinct check nodes, where $\lambda=\Big\lceil\frac{d-2-\lfloor\frac{d-2}{r+1}\rfloor}{r}\Big\rceil$. 
Also, each of the $n-k$ parity nodes is connected to exactly one distinct check node. 
Hence, any $d-1$ variable nodes are connected to at least $d-1$ distinct check nodes and our proposed code has minimum distance $d$.
%Now, we show that any $\phi=d-1$ variable nodes are connected to $d-1$ distinct check nodes which concludes the proof.   
%
%For $\phi= d-1$, the worst case is when all $d-1$ variable nodes are in the same local group because otherwise, they have connection to at least $(d-2)+1=d-1$ check nodes.
%Observe that if $d-1$ variable nodes are in the last local group, they are connected to $d-1$ distinct check nodes. 
%For the case that $d-1$ variable nodes are not in the last local group, we show by contradiction that any $d-1$ variable node is covered by $d-1$ distinct check nodes.
%
%Assuming that all $d-1$ variable nodes are chosen from one of the first $n-k-d+1$ local groups, by contradiction, we assume that $d-1$ variable nodes are connected to one local and $d-3$ global check nodes. 
%In other words, there is a global check node which is not connected to $d-1$ variable nodes in a local group.  
%This contradicts the assumption that each global check node in our proposed LRCs does not cover at most $d-2$ variable nodes.
%